\documentclass[reprint, superscriptaddress,nofootinbib, amsmath,amssymb, aps]{revtex4-2}
\usepackage{xcolor}
\usepackage{braket}
\usepackage{amsmath}
\usepackage{graphicx}
\usepackage{dcolumn}
\usepackage{bm}
\usepackage{hyperref}

\begin{document}


\title{The Second-Level Preheating}

\author{Norma Sidik Risdianto}
\email{norma.risdianto@uin-suka.ac.id}
\affiliation{Research Center for Quantum Physics, National Research and Innovation Agency (BRIN), \\
South Tangerang 15314, Indonesia}
\affiliation{
Department of Physics Education, Universitas Islam Negeri Sunan Kalijaga\\
 Jl. Marsda Adisucipto 55281, Yogyakarta, Indonesia}

\author{Apriadi Salim Adam}
\email{apriadi.salim.adam@brin.go.id}
\affiliation{Research Center for Quantum Physics, National Research and Innovation Agency (BRIN), \\
South Tangerang 15314, Indonesia}
 
\author{Lalu Zamakhsyari}
\email{laluzam@hep.s.kanazawa-u.ac.jp}
\affiliation{Institute Center for Theoretical Physics, Kanazawa University \\
Kanazawa 920-1192, Japan}

\date{\today}

\begin{abstract}
This paper proposes two levels of preheating for the inflaton that is non-minimally coupled with gravity. The first level, later named by the quadratic regime, corresponds to the matter-dominated era and is responsible for draining the inflaton's energy density. The second level, which we will call the quartic regime, corresponds to the radiation-dominated era and is responsible for the reheating of the universe. We investigate the behavior of non-renormalizable higher dimension operators in both the quadratic and quartic regimes. In the quadratic regime, the preheating is also efficient, even though it is less than the lower dimension.
On the other hand, the non-renormalizable higher dimension operators in the quartic regime are extremely inefficient. In our work, we also introduce a simple mechanism controlled by the characteristic momentum $\alpha$ to suppress the particle production during preheating. Additionally, we emphasized the significance of the small momentum of the particles produced during preheating for the abundance of primordial black holes. This result supports the efficient preheating in the quadratic regime. Finally, we evaluate two modes of the reheating temperature, which differ based on the preheating efficiency during the quadratic regime.
\end{abstract}


\maketitle
\section{Introduction}
The new inflation scenario requires a preheating stage to be complete. This stage occurs after inflation ends when the inflaton oscillates and produces particles through non-perturbative processes \cite{dolgov1989particle, traschen1990particle, kofman1994reheating, shtanov1995universe}. These particles can be much heavier than the inflaton's effective mass. Although the preheating stage could complete the inflationary theory, it lacks observational evidence and is highly model-dependent.

Several preheating models have been proposed in recent years. Ref. \cite{kofman1997towards} discusses preheating in the context of a quadratic potential. Most models of inflation with non-minimal coupling \cite{hertzberg, ema2}, Starobinsky inflation \cite{fr, fr2}, and their extensions follow the preheating stages described in these references. Ref. \cite{Greene_structure} presents a model of preheating involving a quartic potential, which has inspired subsequent works such as Refs. \cite{ballesteros2017standard, hashimoto2021inflation}. In Ref. \cite{ema2}, the energy drain due to inflaton oscillation is described as extremely efficient, with depletion occurring in just one zero-crossing of the inflaton. However, the inflaton zero-crossing is milder in two-field inflation models, such as the mixed Higgs-R$^2$ model \cite{he2}.

In our paper, we propose two levels of preheating, termed the first and second levels of preheating. These levels are derived from the Lagrangian with non-minimal coupling, resulting in two stages of preheating. \footnote{Originally, the idea of the second-level preheating was discussed in Ref. \cite{kofman1997towards} with or without backreaction. The preheating with the split quadratic and quartic potential dark matter is also discussed in Ref. \cite{almeida2019hidden}} We define the first-level preheating as the quadratic regime during the matter-dominated era and the second-level preheating as the quartic regime during the later radiation-dominated era. For the second-level preheating to occur, particle production should not be overly efficient during the first-level preheating. In our paper, "efficient" means that the inflaton's energy density could be depleted in just a few zero-crossings. With certain parameter tuning, the inflaton zero-crossing can become so efficient that it drains all of the inflaton's energy in a single crossing. Since we aim for preheating to be less efficient, we calculate such crossing to establish an upper bound on certain parameters. Violating these conditions could prevent the second-level preheating from occurring.

In Ref. \cite{greene2000theory}, it is noted that during preheating, the production of bosonic species may dominate the universe instead of fermions due to Pauli blocking. We analyze this issue during the first-level preheating phase, which occurs during the matter-dominated era. Later, fermions are efficiently produced during the second-level preheating, which takes place during the radiation-dominated era. This distinction will significantly affect the reheating temperature in our model.

This paper is organized as follows: Section \ref{2} discusses the preheating stage which focuses on the first and second-level preheating phases. In section \ref{additional}, we present additional features including preheating for the non-renormalizable higher dimension operator, a brief explanation of the chosen values of the non-minimal coupling $\xi$, the ceased out mechanism, tachyonic preheating, and the production of grand-daughter fields. Later, in section \ref{pbh}, we discuss primordial black holes to obtain constraints on our momentum after inflation. Finally, section \ref{reheatingtemperature} exclusively addresses the reheating temperature. The summary of our research is presented in section \ref{summary}.

\section{The Preheating with inflaton non-minimally coupled with gravity}\label{2}

\subsection{Inflation and Preheating}\label{inflation}
In this subsection, we will consider the action of the inflaton $\phi$ which is non-minimally coupled with gravity as follows\footnote{There is a potential term $\frac{1}{2}m^2_\text{bare}\phi^2$, where $m_\text{bare}$ is the bare mass of inflaton $\phi$. However, it is not relevant to our scenario in this paper. It is assumed to be insignificant},
\begin{equation}\label{actionjordan}
\begin{split}
          S_J&=\int d^4x\sqrt{-g_J}\\
          &\times \left(\frac{1}{2}M_p^2 R+\frac{1}{2}\xi \phi^2 R-\frac{1}{2}g^{\mu\nu}_J\partial_\mu \phi \partial_\nu \phi-\frac{1}{4}\lambda \phi^4\right),  
\end{split}
\end{equation}
where $S_J$ and $R$ are the action and  Ricci scalar in the Jordan frame. $M_p$ is the reduced Planck mass,  $\lambda$ is the quartic coupling of the $\phi$ self-interaction, and $\xi$ is the non-minimal coupling of the inflaton $\phi$ with gravity. This action can be transformed into the Einstein frame by using the following,
\begin{equation}\label{transformation}
      g_{\mu\nu E}=\Omega^2 g_{\mu\nu J}, \hspace{1cm}\Omega^2\equiv 1+\frac{\xi \phi^2}{M_p^2}\equiv e^{\sqrt{\frac{2}{3}}\frac{\chi}{M_p}}.
\end{equation}
The subscript $E$ denotes the Einstein frame. Then, the action in the Einstein frame is given by
\begin{equation}\label{actioneinstein}
    S_E=\int d^4x\sqrt{-g_E}\left(
    \frac{1}{2}M_p^2 {R}_E-\frac{1}{2}g^{\mu\nu}_E\partial_\mu \chi \partial_\nu \chi -U(\chi)\right), 
\end{equation}
where ${R}_E=R\Omega^{-2}-6\Omega^{-3}\square\Omega$. In the rest of the paper, the subscript $E$ will be omitted for simplicity. Note that we have omitted non-canonical kinetic terms due to the scale of inflation in the above expression. 

Next, the potential in Eq.\eqref{actioneinstein} has the following form
\begin{equation}\label{uchi}
         U(\chi)=\frac{\lambda}{4\Omega^2}\phi^4=\frac{\lambda}{4\xi^2}M_p^4\left( 1-e^{-\sqrt{\frac{2}{3}}\frac{\chi}{M_p}}\right)^2.
\end{equation}
The power spectrum $P_s$ is written as follows,
\begin{equation}\label{Ps}
    P_s=\frac{1}{12\pi^2}\frac{U^3}{M_p^6 (\partial U/\partial \chi)^2}\Bigg|_{\text{ini}}. 
\end{equation}
If we use $P_s=2.101 \times 10^{-9}$ at $\kappa_*=0.05$ Mpc$^{-1}$ \cite{cmb} and number of e-folds $N=56$, we obtain the constraint of the ratio $\frac{\xi^2}{\lambda}$ during inflation as follows,
\begin{equation}\label{constraintcmb}
    \frac{\xi^2}{\lambda}\simeq 2.1 \times 10^9.
\end{equation}
In general, both $\xi$ and $\lambda$ are not fixed and depend on the running coupling of their energy scale. This means the constraint from the CMB in Eq. \eqref{constraintcmb} may not apply at lower energy levels. Thus, after the end of inflation, $\xi$ and $\lambda$ are no longer constrained by CMB. However, studying the running coupling is beyond the scope of this paper.

Below we define two regimes of the preheating stages (after inflation) based on the potential as follows,
\begin{equation}\label{potentialsplit}
    U \simeq
    \begin{cases}
    \frac{1}{2}\left(\frac{\lambda }{3\xi^2}M_p^2 \right) \chi^2         & \text{if } \Tilde{\chi}>\Tilde{\chi}_\text{crit}\\
    \frac{1}{4}\lambda \chi^4 & \text{if } \Tilde{\chi}< \Tilde{\chi}_\text{crit}
    \end{cases},
\end{equation}
where $\Tilde{\chi}_\text{crit}\approx \Tilde{\phi}_\text{crit}\approx\frac{M_p}{\xi}$. The tilde refers to the amplitude of the fields. This critical value corresponds to the field value when the amplitude of the scalaron $\chi$ in the Einstein frame coincides with the amplitude of $\phi$ in the Jordan frame. 
In Eq. \eqref{potentialsplit}, the condition of $\Tilde{\chi}>\Tilde{\chi}_\text{crit}$ corresponds to the quadratic regime, as indicated by the $\chi^2$ on the potential. Conversely, the condition $\Tilde{\chi}<\Tilde{\chi}_\text{crit}$ corresponds to the quartic regime, indicated by the $\chi^4$ term in the potential.

\subsection{The First-Level Preheating, The Quadratic Regime}\label{quadraticregime}

In the following, we will investigate the feature of the inflaton in the quadratic regime. This regime is defined when the field value of the inflaton is $\Tilde{\chi}<  M_p $, as stated in the previous subsection. At this time, the inflaton oscillates around zero and the slow-roll conditions are violated ($\epsilon=1$). After the end of inflation, the potential in Eq. \eqref{potentialsplit} can be rewritten as follows,
\begin{equation}\label{m}
    U(\chi) \equiv\frac{1}{2}m^2\chi^2,
\end{equation}
where $m^2=\frac{\lambda}{3\xi^2}M_p^2$ corresponds to the effective inflaton mass during preheating and is strongly constrained by Eq. \eqref{constraintcmb}.
The equation of motion obtained from Eq. \eqref{actioneinstein} can be written as 
\begin{equation}\label{squarechi}
    \square \chi - \frac{dU}{d\chi}=0.
\end{equation}

If we write the field $\chi$  by using the Heisenberg representation as
\begin{equation}\label{heisenberg}
   \chi(\textbf{x},t)=\int \frac{d^3k}{(2\pi)^{\frac{3}{2}}}\left(\hat{a}_k \chi_{k}(t)e^{-i\textbf{k}\cdot \textbf{x}} +\hat{a}^\dagger_k \chi^*_{k}(t)e^{i\textbf{k}\cdot \textbf{x}}\right).
\end{equation}
where $\hat{a}_k$ and $\hat{a}^\dagger_k$ corresponds to the annihilation and creation operators, 
we obtain
\begin{equation}\label{selfphi}
\Ddot{\chi}_k+3H\Dot{\chi}_k+\left(\frac{k^2}{a^2}+m^2\right)\chi_k=0.
\end{equation}
Note that the dot represents the derivative with respect to the physical time $t$ and $k$ is the momentum. During this time, the universe enters the matter-dominated era due to the nature of the potential $U(\chi)$. It is obvious from Eq. \eqref{selfphi} that the resonance of self-$\chi$ production is conformally suppressed \cite{kofman1997towards}. 

Next, we add another Lagrangian to Eq. \eqref{actionjordan}, which has the form as
\begin{equation}\label{interaction}
    -g^{\mu\nu}_J\partial_\mu\psi\partial_\nu\psi-\frac{1}{4}g\psi^2 \phi^2,
\end{equation}
where $g$ is the real-valued coupling and $\psi$ is another scalar field. The interaction terms can be provided in the Einstein frame up to the leading order as
\begin{equation}\label{trilinear}
    -\frac{1}{2}g'\psi^2|\chi|,
\end{equation}
where 
\begin{equation}\label{gprime}
    g'\equiv g\frac{M_p}{\xi\sqrt{6}}.
\end{equation}

Below, we will calculate the particle production of the $\psi$ field by the resonance of the inflaton field $\chi$. 
By  using the Heisenberg representation similar to Eq. \eqref{heisenberg}, we can redefine $\psi$ and obtain the equation of motion as

\begin{equation}\label{quadraticpsi}
    \frac{d^2{\psi_k}}{dt^2} +3H\Dot{\psi}_k+\left(\frac{k^2_\psi}{a^2}+{g'}|\Tilde{\chi}\sin(m t)|\right){\psi_k}=0,
\end{equation}
where we have used
\begin{equation}\label{chitilde}
    \chi=\Tilde{\chi}\sin(m t).
\end{equation}
This equation of motion is similar to that in Ref. \cite{dufaux2006preheating}. The absolute value is due to the consequences of Eq. \eqref{transformation}.

By adopting the procedure on Ref. \cite{Bezrukovinitials}, we can obtain the equation of motion as
\begin{equation}\label{analyticpsiquadratic}
    \frac{d^2\psi_k}{dq^2}+(p^2+q)\psi_k=0,
\end{equation}
where 
\begin{equation}\label{pqK}
    p^2=\frac{k^2_\psi}{K^2}, \hspace{1cm} q=Kt, \hspace{1cm}K=\left[g' \Tilde{\chi}m \right]^{1/3},
\end{equation}
and we have approximated $\Tilde{\chi}\sin(m t)\simeq \Tilde{\chi}m t$ and $a=1$ on the zero-crossing. For $p^2>q$, the $\psi_k$ particle production is mostly inefficient. Greater $q$ impacts the larger particle production. For $p^2\ll q$, the particle production is extremely efficient and can be estimated analytically as
\begin{equation}\label{energyproductionpsi}
    \Tilde{\rho}_\psi=\int^\infty_0 \frac{d^3k_\psi}{(2\pi)^3}e^{-\pi p^2}\sqrt{g'\Tilde{\chi}}.
\end{equation}

The particle produced by inflaton oscillation is mostly heavy and non-relativistic during this matter-dominated era. Hence, the contribution of the momentum $k_\psi$ is neglected.
In section \ref{pbh}, we will show that the large momentum is discouraged for appropriate primordial black hole abundance. By solving  Eq. \eqref{energyproductionpsi}, we can obtain the energy produced during the (first) single-crossing to be

\begin{equation}\label{deltarhopsi}
\frac{1}{4\pi^3} \left(g'\Tilde{\chi}\right)^{3/2}m\simeq \left(\frac{g}{\xi}\right)^{3/2} 2.65 \times 10^{-8} M_p^4,
\end{equation}
where we have used $\Tilde{\chi}\simeq  \hspace{0.5mm}M_p$ that corresponds to the field value during the start of the preheating stage.
Comparing the inflaton energy to be $\frac{1}{2}m^2 \chi^2\simeq 8 \times 10^{-11} M_p^4 $ with our result on Eq. \eqref{deltarhopsi}, the inflaton's energy could  potentially be drained by the first single-crossing if we take the value

\begin{equation}\label{singlecrossingmatterdominated}
    \frac{g}{\xi} \gtrsim 0.02.
\end{equation}
In realistic models such as Higgs inflation \cite{bezrukovstandard} (See also \cite{garcia2009preheating,repond2016combined}), producing gauge bosons through parametric resonance is less efficient due to large $\xi$ and small gauge boson couplings. Consequently, the violation of Eq. \eqref{singlecrossingmatterdominated} results in less efficient preheating. It is important to note that, according to the criterion in Eq. \eqref{singlecrossingmatterdominated}, efficient preheating is favored by small $\xi$ and large $g$. Since $g$ is typically smaller than unity, values of $\xi \gtrsim \mathcal{O}(100)$ are not favorable for efficient preheating. However, in the context of Higgs inflation, the decay of gauge bosons is nearly instantaneous, which favors a high reheating temperature \cite{Bezrukovinitials}.

We can see that the second-level preheating exists if Eq. \eqref{singlecrossingmatterdominated} is not fulfilled.
Thus, the criterion for second-level preheating can only be valid if we rewrite it to be

\begin{equation}\label{2ndpreheatingsyarat}
    \frac{g}{\xi} < 0.02.
\end{equation}
In other words, second-level preheating is favored by the inflation model with smaller $g$ and larger $\xi$.

For the numerical calculation of Eq. \eqref{quadraticpsi}, we need to define 
\begin{equation}\label{az}
   mt=2z-\pi/2, \hspace{0.5cm}A=\frac{4k_\psi^2}{m^2}, \hspace{0.2cm}\text{and}\hspace{0.2cm}Q=\frac{2gM_p}{m^2\xi\sqrt{6}}\Tilde{\chi}.
\end{equation}
Thus, neglecting the expanding universe effect ($a=1$), we obtain
\begin{equation}\label{a2q2}
    \frac{d^2\psi_k}{dz^2}+(A-2|Q \cos (2z)|)\psi_k=0,
\end{equation}
from which it belongs to the Mathieu equation but with the absolute value. Note that we used $\Tilde{\chi}=M_p$, corresponding to the field value during the end of inflation. Also, it is implied that $Q$ is always $\gg 1$ without fine-tuning and favoring the heavy non-relativistic particle productions.

\begin{figure}[t]
\centering
\includegraphics[width=8cm]{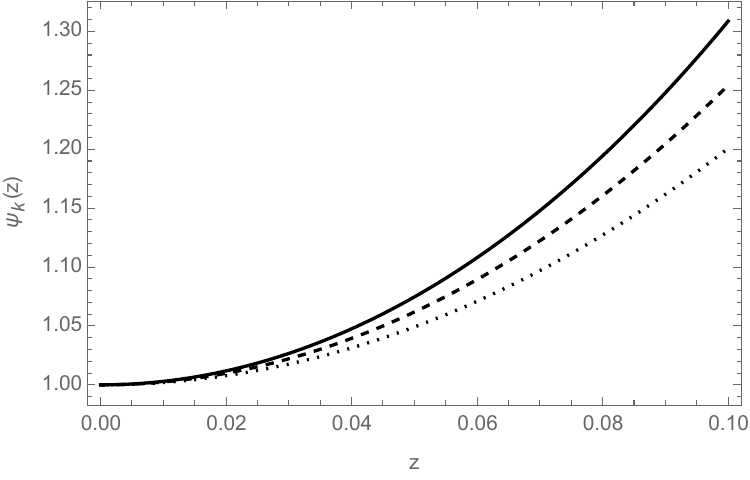}
\caption{The growth of $\psi_k$. We used fixed $A=1$, $Q=20$ (dotted), $Q=25$ (dashed), and $Q=30$ (solid line) for this numerical calculation in our sample. Note that $z$ is in a unit of $\pi/2$ and we neglected the expansion of the universe ($a=1$). This figure shows the growth of the mode $\psi_k$ due to resonance. For the numerical calculation, we take $\psi_k(0)=1$ and $\frac{d\psi_k}{dz}\big|_{z=0}=0$.}
\label{quadraticfigure}
\end{figure}
\begin{figure}
    \centering
\includegraphics[width=8cm]{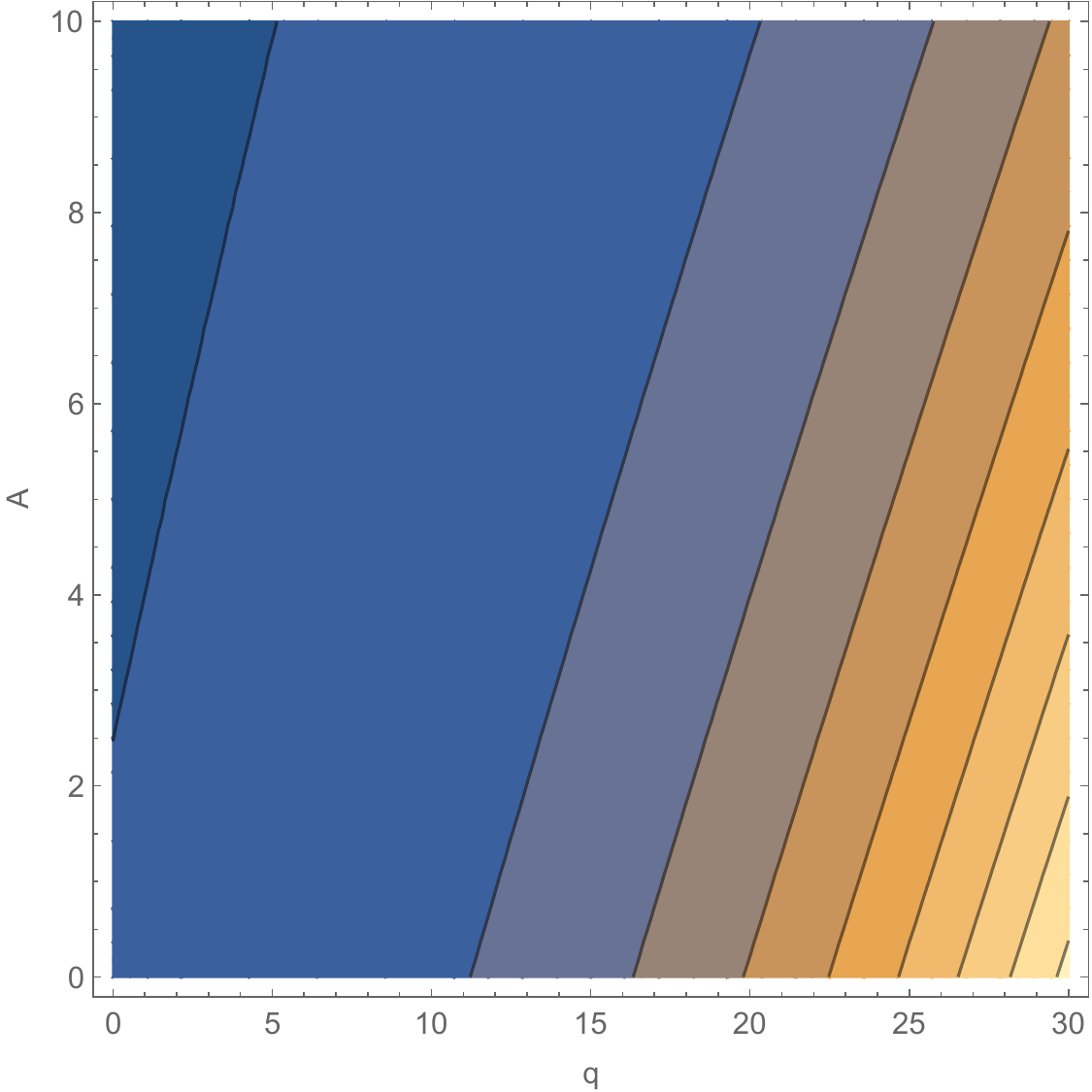}
    \caption{The Mathieu-like instability chart with corresponding values of $A$ and $Q$ and evaluated at $z=1$. The colors of the figure's legends show the characteristic exponent $\mu_p$. The brightest image refers to $\mu_p=1.6$. The darker color decrease the $\Delta \mu_p$ by $0.2$ }
    \label{mathieustabilityfigure}
\end{figure}

We show the numerical calculation of Eq. \eqref{a2q2} in Fig. \ref{quadraticfigure}. The varied $A$ and $Q$ parameters obey the Mathieu-like instability chart in Fig. \ref{mathieustabilityfigure}, in which the colored legends correspond to the characteristic exponent $\mu_p$ that can be written analytically as,
\begin{equation}\label{characteristicexponent}
    \mu_p=\frac{1}{2\pi}\ln\left(1+2e^{-\pi p^2}-2\sin\theta_p e^{-\frac{\pi}{2} p^2}\sqrt{1+2e^{-\pi p^2}}\right),
\end{equation}
where $\theta_p$ is a phase.
As $Q$ is always much larger than $A$ especially during the first zero crossing,  the particle production can be estimated analytically as in Eq. \eqref{energyproductionpsi}.

\subsection{The Second-Level Preheating, The Quartic Regime}\label{quarticregime}

During the preheating stage, the field value of the $\chi$ drops each crossing. After some point where $\Tilde{\chi}\approx \Tilde{\phi} \approx M_p/\xi$, the action in the Einstein frame (Eq. \eqref{actioneinstein}) coincides with the Jordan frame (Eq. \eqref{actionjordan}), making them nearly indistinguishable \cite{lebedev2023dark} and the quartic regime, as the second-level preheating, is started\footnote{We adopt this mechanism in our earlier work, see Ref. \cite{risdianto2022inflation}}.  
One should have a note since it is still debatable which frame is more physical than the other. It is advisable in Ref. \cite{capozziello2010physical} to consider only one frame when discussing the whole inflationary scenario. Consequently, even if we seem to work in the Jordan frame for this second-level preheating, we still work in the Einstein frame. During this time $\Omega^2= 1$, the potential of Eq. \eqref{uchi} can be written as $\frac{1}{4}\lambda \chi^4$ (See Eq. \eqref{potentialsplit}). However, as $\phi$ is coincided with $\chi$ during this regime, we can write the potential as

\begin{equation}\label{potentialquartic}
    \frac{1}{4}\lambda \phi^4,
\end{equation}
where the quartic regimes could be easily distinguished from the quadratic regimes (as it uses $\chi$ as the inflaton).
The equation of motion of inflaton $\phi$ which is derived from Eq. \eqref{actionjordan} turns out to be
\begin{equation}
    \frac{d^2\phi}{dt^2}+3H\Dot{\phi}+3\lambda\phi^3=0.
\end{equation}
If we used the definition of the conformal field 
\begin{equation}\label{varphi}
    \varphi=a\phi=a\Tilde{\phi}f=\Tilde{\varphi}f
\end{equation}
and conformal time $d\tau=\sqrt{\lambda}\Tilde{\phi}\frac{dt}{a}$, we obtain 
\begin{equation}\label{phi3}
    \frac{d^2f}{d\tau^2}+f^3=0.
\end{equation}
The solution of Eq. \eqref{phi3} is \cite{Greene_structure}
\begin{equation}
    f=\text{cn}\left(\tau,\frac{1}{\sqrt{2}} \right),
\end{equation}
where $\text{cn}\left(\tau,1/\sqrt{2}\right)$ corresponds to the Jacobi elliptic function.
With this result, we can calculate the $\psi$ production by using the equation of motion from Eq. \eqref{interaction} by considering the quantum fluctuation of $\psi$-field in the Heisenberg picture via
\begin{equation}
    \frac{d^2\psi_k}{dt^2}+3H\Dot{\psi}_k+\left(\frac{k^2_\psi}{a^2}+\frac{1}{2}g\phi^2 \right)\psi_k=0.
\end{equation}
If we define 
\begin{equation}\label{psibesar}
    \Psi=a\psi_k, \hspace{1cm}  {k}_\Psi=\frac{k_\psi}{a\sqrt{\lambda} \Tilde{\phi}},
\end{equation}
and using the conformal time $\tau$, we obtain
\begin{equation}\label{psiquarticconformal}
    \frac{d^2\Psi}{d\tau}+\left(k_\Psi^2+\frac{g}{2\lambda}\text{cn}^2\left( \tau, \frac{1}{\sqrt{2}}\right)\right)\Psi=0.
\end{equation}

\begin{figure}
    \centering
    \includegraphics[width=8.5cm]{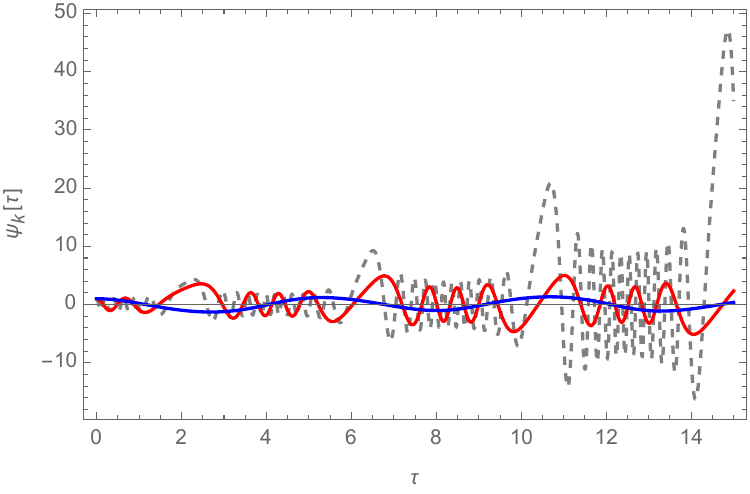}
    \caption{The production growth of  $\psi_k$ as a function of the conformal time $\tau$ in the quartic regime. In this plot, we have used fixed $k_\Psi=1$ and $g/2\lambda=1000$ (dashed line), $g/2\lambda=100$ (red line), and $g/2\lambda=1$ (blue line). This numerical calculation is based on Eq.\eqref{psiquarticconformal}. For the numerical calculation, we take $\Psi(0)=1$ and $\frac{d\Psi}{d\tau}\big|_{\tau=0}=0$.}
    \label{quarticfigure}
\end{figure}

\begin{figure}
    \centering
    \includegraphics[width=8.5cm]{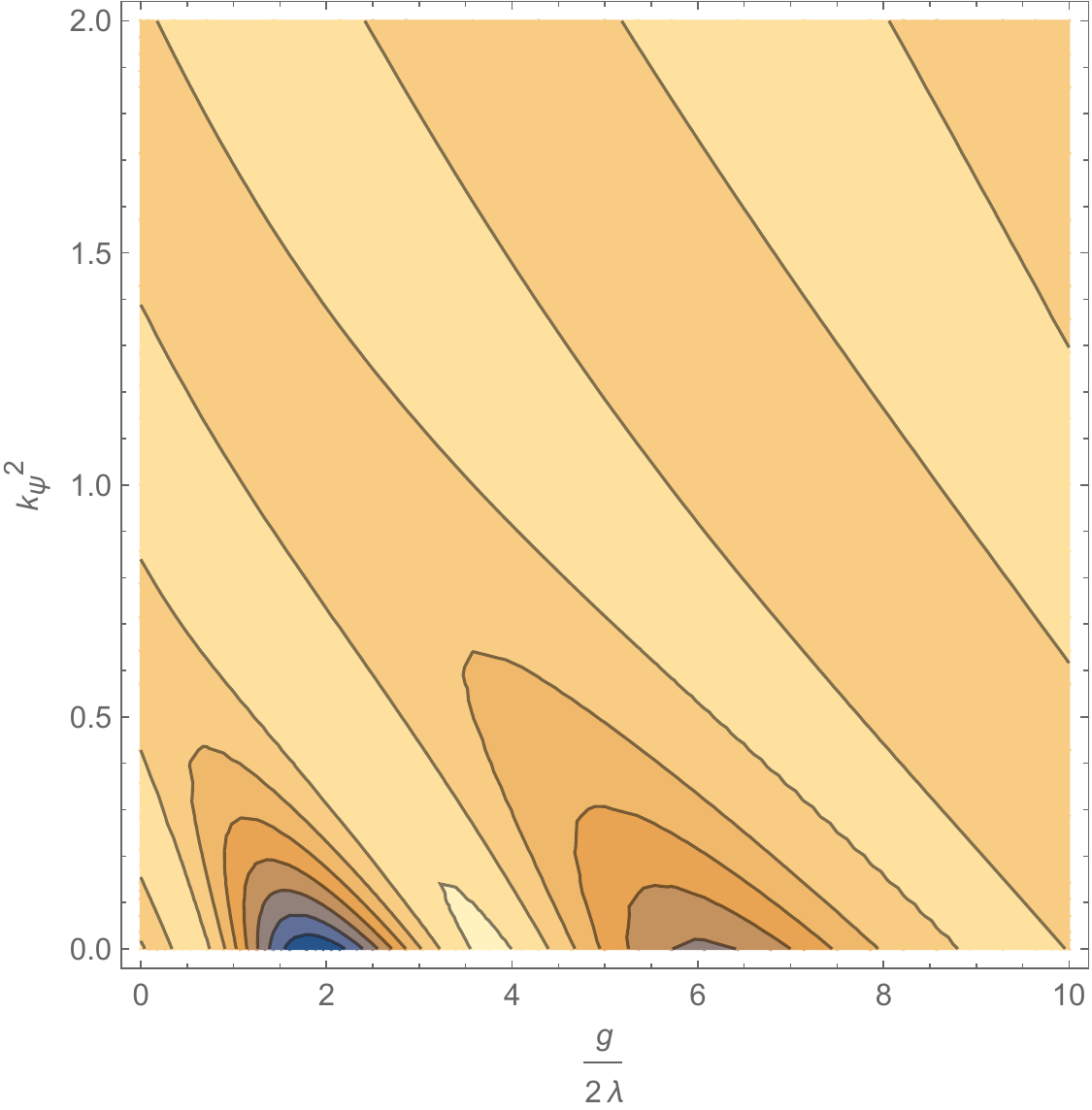}
    \caption{The Lame instability plot with varied $k_\psi^2$ and $g/2\lambda$ evaluated at $\tau=12$. The darker the plot depicts the greater the characteristic exponent $\mu_p$ related to the growth of $\psi_k$. The darkest region corresponds to the characteristic exponent $\mu_p = 0.12$. Every brighter shade shows a decrease of $\Delta \mu_p=0.02$. The characteristic exponent in this quartic regime resemble with $\mu_p$ (of the quadratic regime) in Eq. \eqref{characteristicexponent} with $p$ interchanged with $\frac{k_\psi}{\sqrt{g/2\lambda}}$}
    \label{lamestability}
\end{figure}

The numerical calculation of Eq. \eqref{psiquarticconformal} is depicted in Fig. \ref{quarticfigure}. In this figure, we used $g/2\lambda=1000, 100$, and $1$ for the fixed value of $k_\psi=1$. These three values correspond to the three regimes on the behavior of the Lame equation depicted in Eq. \eqref{psiquarticconformal}. Generally, the greater $g/2\lambda$ corresponds to the larger $\psi_k$ growth. For $k_\psi \sim  g/2\lambda$, the $\psi_k$ growth is affected by the instability chart for the Lame equation (See Fig. \ref{lamestability}). In our paper, we primarily consider effective preheating as the upper bound. In that case, it is favored by the $k_\psi \ll  g/2\lambda$. This condition can be approximated analytically to obtain the constrained upper bound as we show it shortly.

In the following, we can approximate the conformal energy density produced by the resonance during this quartic regime for a single crossing as 
\begin{equation}\label{quarticenergy}
\delta \Tilde{\rho}_\Psi\approx\int^\infty_0\frac{d^3{k}_\Psi}{(2\pi)^3}\exp\left(-\pi\frac{k_\Psi^2}{g/4\lambda}\right)\sqrt{\frac{g}{2\lambda}}\simeq 0.044 \hspace{1mm}\frac{g}{\lambda},
\end{equation}
which corresponds to the physical energy density (neglecting the expansion of the universe) as
\begin{equation}
    \delta \rho_\psi =\lambda^2\Tilde{\phi}^4 \times  0.044 \hspace{1mm}\frac{g}{\lambda} \simeq 0.01 g \times \frac{1}{4}\lambda \Tilde{\phi}^4.
\end{equation}
With $g$ should not be more than unity, the second-level preheating could not efficiently drain the inflaton energy. Thus, it does not violate the late stage of the preheating which favors the narrow resonance condition. Our simple result is independent of $\xi$ as it is supposed to be. 

\section{The Additional Features}\label{additional}
\subsection{The non-renormalizable higher dimension operator preheating}\label{nonrenormalizable}

In the preheating, we have already calculated the trilinear interaction (first-level preheating) and the four-legs interaction (second-level preheating). However, we suggest that the possibility of a higher dimensional term also plays a significant role in the preheating stage (See Ref. \cite{dufaux2006preheating}).

\subsubsection{Quadratic regime}
Straightforwardly,  we start with
the potential on the dimension-n operator as,
\begin{equation}\label{higher-order}
    -\frac{y_n}{M_p^{n-2}}\psi^2\phi^n,
\end{equation}
where $y_n$ is the higher dimensional couplings (we assume it is real-valued) and $n$ is an integer. For higher orders, we start $n=3$ for the dimension-$5$ operator. Typically,  the denominator of Eq.\eqref{higher-order} contains some UV cut-off with the same order as Planck mass $M_p$. Thus, we used $M_p$ for simplification.

\begin{figure}
    \centering
    \includegraphics[width=8.5cm]{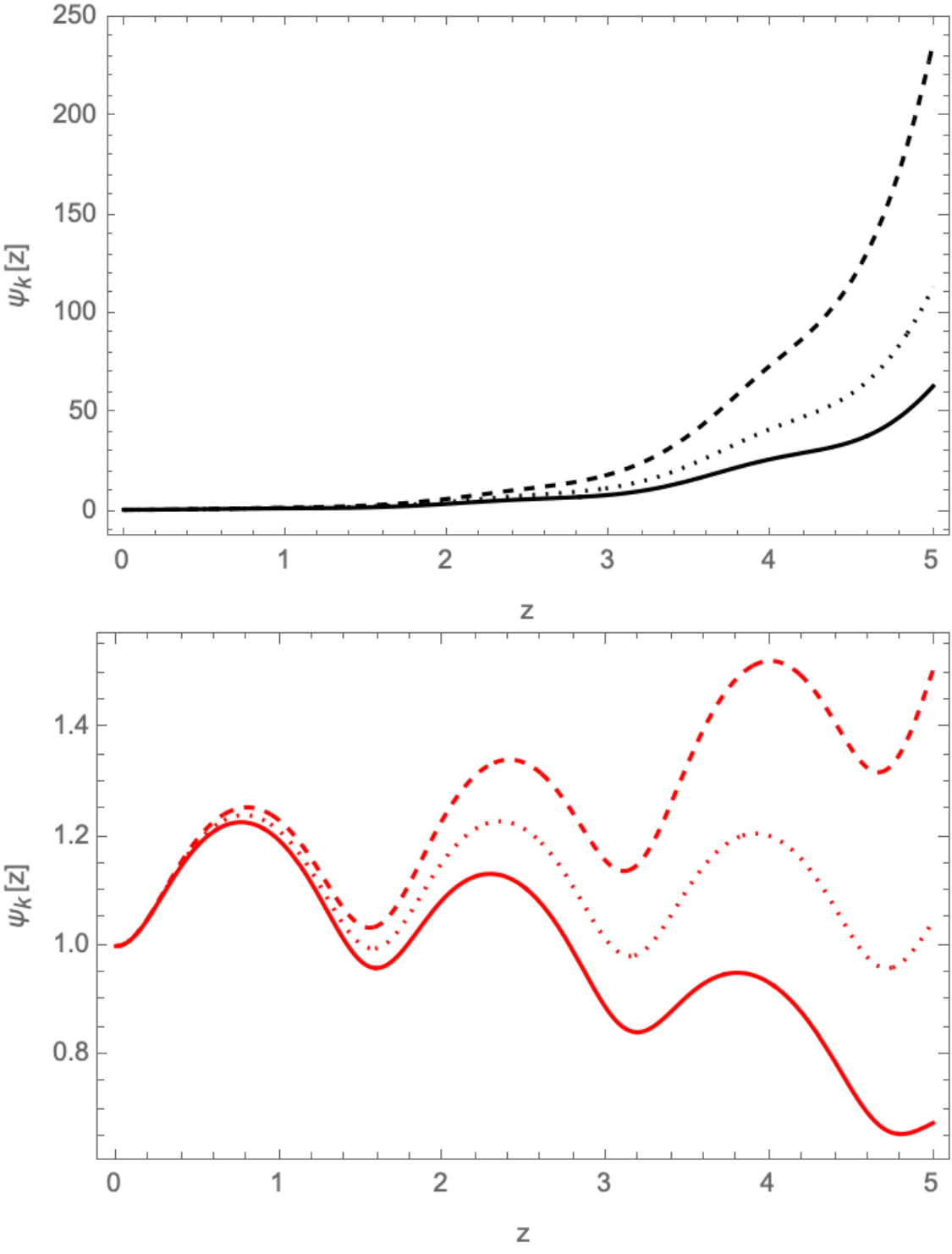}
    \caption{The upper panel shows the growth of $\psi_k$ with $n=3$, $4$, and $5$ while the lower panel with $n=15$, $16$, and $17$. In this figure, we used $A=1$ and $q_n=2$.}
    \label{higherdimquadraticfigure}
    \end{figure}

By using the Weyl transformation in Eq. \eqref{transformation}, we can write Eq. \eqref{higher-order}  in the Einstein frame as (up to the leading order of $\chi$)
\begin{equation}\label{interactioneinstein}
-\left(\frac{2}{3}\right)^{\frac{n}{4}}\frac{y_n}{\xi^{\frac{n}{2}}M_p^{\frac{n}{2}-2}}\psi^2|\chi|^{\frac{n}{2}}.
\end{equation}
The half-integer on the power of $\chi$ is strange for conventional field theory. However, it is allowed on a low-energy effective description for string-derived inflationary
theories such as axion monodromy \cite{silversteinmonodromy} (See also Ref. \cite{cline2022asymmetric}).
The equation of motion of the $\psi$ field for the higher dimension in the Heisenberg picture can be  approximated as
\begin{equation}\label{eomhigh}
    \frac{d^2\psi_k}{dt^2}+3H\Dot{\psi}_k+\left(\frac{k^2_\psi}{a^2}+\frac{2\left(2/3\right)^{\frac{n}{4}}}{\xi^{\frac{n}{2}}M_p^{\frac{n}{2}-2}}y_n|\chi|^{\frac{n}{2}}\right)\psi_k=0.
\end{equation}
Furthermore, if we define  
\begin{equation}\label{qn}
    q_n=\frac{4(2/3)^{\frac{n}{4}}y_n\Tilde{\chi}^{\frac{n}{2}}}{m^2\xi^{\frac{n}{2}}M_p^{\frac{n}{2}-2}}
\end{equation}
and using the definition of $A$ and $z$ in Eq. \eqref{az}, we finally obtain the $n$-dependent Mathieu-like equation as
\begin{equation}\label{mathieu-n}
    \frac{d^2\psi_k}{dz^2}+\left(A-2|q_n (\cos(2z))|^{\frac{n}{2}} \right)\psi_k=0.
\end{equation}
In FIG. \ref{higherdimquadraticfigure}, We show the numerical calculation of Eq. \eqref{mathieu-n} with $n=3$, $4$, and $5$ in the upper panel while $n=15$, $16$, and $17$ in the lower panel.
It shown that the higher $n$ implied that the preheating is more inefficient. In addition, even though $q_n$ is guaranteed to be $\gg 1$, the preheating in this higher-order quadratic regime is less likely to be more efficient than the lower order (See FIG. \ref{quadraticfigure}).

As a demonstration, we consider $n=4$ for analytical calculation. The solution is straightforward and helps to clarify our numerical results, as shown in the upper panels of FIG. \ref{higherdimquadraticfigure}. Specifically, particle production from Eq. \eqref{eomhigh} with $n=4$ simplifies to
\begin{equation}\label{psidim6}
    \frac{d^2\psi_k}{dt^2}+\left(k^2_\psi+\frac{4y_4}{3\xi^2}\Tilde{\chi}^2\sin^2(m t)\right) \psi_k=0,
\end{equation}
where we used Eq. \eqref{chitilde} and $a=1$. At the small $mt$, we obtain
\begin{equation}\label{psidim6x}
    \frac{d^2\psi_k}{dt^2}+\left(k^2_\psi+\frac{4y_4}{3\xi^2}\Tilde{\chi}^2m^2 t^2\right) \psi_k=0.
\end{equation}
If we define the dimensionless momentum $k_4=\frac{k_\psi}{\sqrt{\Tilde{\chi}m}}$ and dimensionless time $q_4=\sqrt{\Tilde{\chi}m}t$ we obtain
\begin{equation}
    \frac{d^2\psi_k}{dq_4^2}+\left(k^2_4+\frac{4y_4}{3\xi^2}q_4^2\right) \psi_k=0.
\end{equation}
For $\frac{4y_4}{3\xi^2}q_4^2\gg k_4^2$, the conformal energy density by the single crossing can be analytically approximated as  
\begin{equation}
\delta \Bar{\rho}_4 =\int^\infty_0 \frac{d^3k_4}{(2\pi)^3}\exp\left(-\pi \frac{k_4^2}{4y/3\xi^2}\right)\sqrt{\frac{4y_4}{3\xi^2}}=\frac{1}{3\pi^3} \frac{y}{\xi^2},
\end{equation}
In addition, it corresponds to the physical density
\begin{equation}
    \delta\rho_4 = m^2\Tilde{\chi}^2 \cdot \delta\Bar{\rho}_4 = \left(\frac{2}{3\pi^3} \frac{y_4}{\xi^2}\right)\cdot \frac{1}{2}m^2 \Tilde{\chi}^2.
\end{equation}
It means that to drain the inflaton energy by a single crossing for this dimension-6 operator, we need a constraint as
\begin{equation}\label{dim6quadratic}
    \frac{2}{3\pi^3} \frac{y_4}{\xi^2}\gtrsim 1.
\end{equation}
It is obvious, as $y_4$ (or $y_n$) is always in order of unity (or smaller), even if we put $\xi$ to be in the order of unity, the criterion Eq. \eqref{dim6quadratic} is still impossible to fulfill. 

\subsubsection{Quartic regime}

\begin{figure}[t]
\centering
\includegraphics[width=8cm]{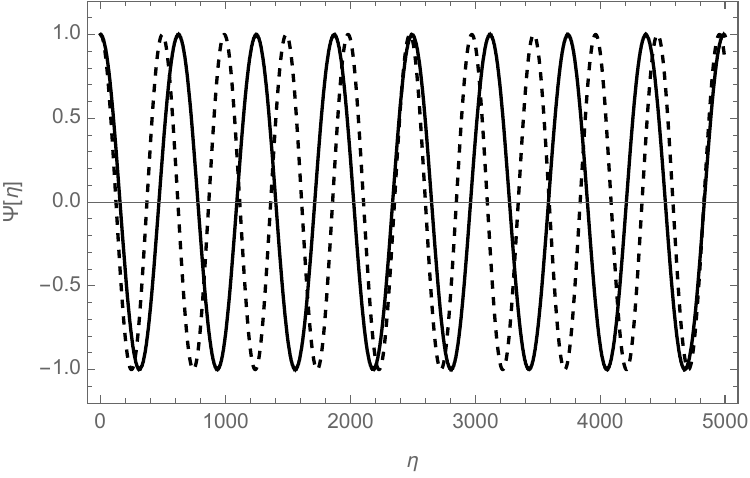}
\caption{ The preheating in the quartic regime for the dimension-n operator in the conformal mode. We used $n=3$ (solid-black) and $n=4$ (dashed) as representations since higher $n$ shows similar behavior but only differ in their phases. We used $k_\psi=0.01$, $y_n=1$, $\xi=10$, $a=1$, $\Psi(0)=1$, $\frac{d\Psi}{d\eta}\big|_{\eta=0}=0$, and $\Tilde{\phi}=M_p/\xi$ for this numerical calculation. We also varied the parameters ($k_\psi$, $y_n$, and $\xi$) but they do not show significant results.}
\label{dim5quarticfigure}
\end{figure}

When the field value of the inflaton drops to $Mp/\xi$, the second-level preheating occurs. The equation of motion in the conformal mode of the dimension-$n$ operator in the Heisenberg picture can be written by 
\begin{equation}\label{dim-n-quartic}
    \frac{d^2\Psi}{d\eta^2}+\left(k_\psi^2+\frac{2y_n a^2}{M_p^{n-2}}\Tilde{\phi}^n\text{cn}^n\left(\eta, \frac{1}{\sqrt{2}}\right) \right)\Psi=0,
\end{equation}
where it has been derived from the Lagrangian with potential \eqref{higher-order}. We also use the definition of $\Psi$ on Eq. \eqref{psibesar}, $d\eta=\frac{dt}{a}$,  and Eq. \eqref{varphi}. The numerical calculation can be seen in FIG. \ref{dim5quarticfigure}. In that figure, it is clear that the preheating is extremely inefficient compared to the lower-dimension operators. 

\subsection{On $0<\xi<1$}\label{xiless1}
In this paper, the first (quadratic) and the second (quartic) preheating are split strongly by the value of $\xi$. For example, we consider the end of inflation to be marked when $\chi= M_p$. If we take $\xi=10$, the quartic regime begins when the field value reaches approximately  $0.1M_p$ based on Eq. \eqref{potentialsplit}. However, for $\xi=1000$, the quartic regime starts at a much smaller field value, around $\sim 0.001 M_p$. 
A problem arises if we take $0<\xi<1$. According to the definition in Eq. \eqref{potentialsplit}, the starting point of the quartic regime would occur at a field value larger than the end of inflation, which is not possible.

To answer this issue, let us reconsider when $\xi=10$. In this scenario, the starting point of the quartic regime is at one-tenth of the inflaton field value at the end of inflation. This implies that the quadratic regime spans field values from approximately $M_p$ down to $0.1 M_p$. If we decrease $\xi$, the duration of the quadratic regime shortens, and the quartic regime begins at a higher energy. Following this logic, when $\xi = 1$, the quadratic regime is effectively skipped. It means, that just after the end of the inflation, it is straightforwardly continuing into the quartic regime \cite{hashimoto2021inflation}. 
This condition should be applicable for smaller $\xi$. 

\subsection{The ceased-out mechanism}\label{cease-out}

It is important to note that in FIG. \ref{quadraticfigure}, the growth of $\psi_k$ is enormous.  Thus, if the crossing is so strong, the preheating will end in the first-level preheating with no remaining inflaton's energy for the second-level preheating.

In the numerical study of the preheating, we treat momentum $k_\psi$ as a fixed value. The large $k_\psi$ corresponds to the condition when the produced particles are relativistic. On the contrary, the smaller $k_\psi$ refers to the heavy-non relativistic particle production. Realistically, at the beginning of the preheating stage, the particle produced during the early crossing should be stronger than the next crossing since the inflaton's oscillation induces it. 

Thus, it is more convenient to say that $k_\psi$ should be time-dependent. In this case, the function of $k_\psi(t)$ should behave like a cease-out mechanism for $\psi$. To straightforwardly explain this matter, let us define  the momentum evolution $k_\psi(t)$ as
\begin{equation}\label{kt}
    k_\psi(t)=k_0 e^{\alpha t},
\end{equation}
where $\alpha$ refers to a constant.
It is more convenient to dub this parameter as the \textit{characteristic momentum}.
Eq. \eqref{kt} shows that it exponentially grows with respect to $t$.

In the preheating scenario. Originally, the fixed momentum and the coupling strength between the produced and inflaton fields solely give the fate of the created particle density \cite{kofman1997towards, Greene_structure}. However, in this paper, 
we introduce the characteristic momentum $\alpha$ as the selector of the produced fields. In this case, the momentum $k_\psi$ becomes time-dependent.

\begin{figure}[t]
    \centering
\includegraphics[width=8.5cm]{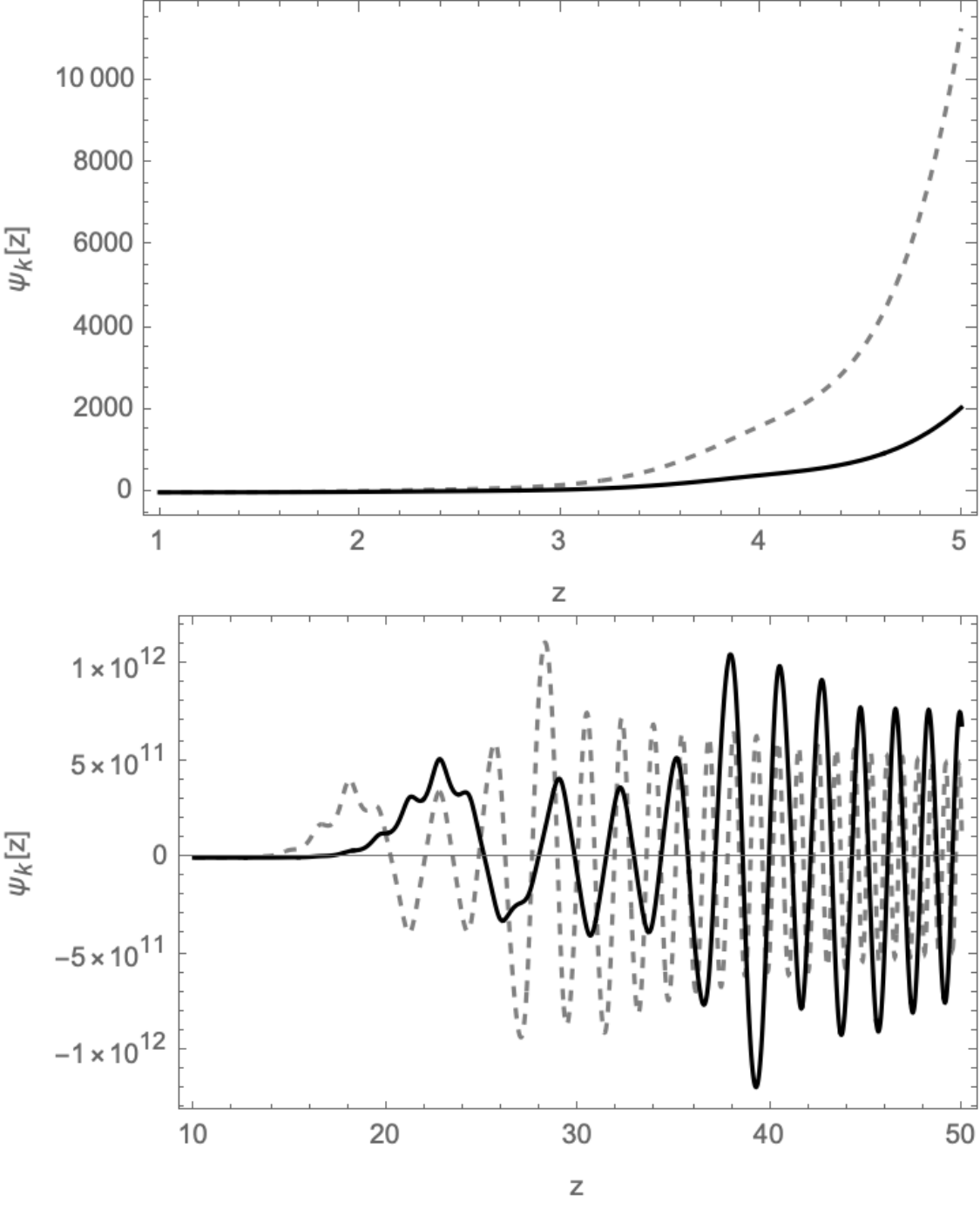}
    \caption{The upper figure corresponds to the $\psi_1$ (dashed-line) and $\psi_2$ (solid-line) growth. It is shown that due to the larger coupling, $\psi_1$ overwhelmed the $\psi_2$ production. In the lower figure, the $\psi_2$ production is higher than $\psi_1$ due to a smaller characteristic momentum ($\alpha_1>\alpha_2$). The $z$ in unit of $\pi/2$. In this figure, we used $\alpha_1=0.045$ and $\alpha_2=0.03$. We also set $\Tilde{A}_1=\Tilde{A}_2=1$, $Q_1=4$, and $Q_2=3$. It is shown that when $z\approx 38$, $\psi_2$ surpasses the growth of $\psi_1$.}
    \label{charmomentum}
\end{figure}

To be more realistic, we can consider 2 fields: $\psi_1$ and $\psi_2$. Both coupled with inflaton $\phi$ via interactions
\begin{equation}
        \frac{1}{4}g_1\psi_1^2 \phi^2 \hspace{0.5cm}\text{and}\hspace{0.5cm}\frac{1}{4}g_2\psi_2^2 \phi^2.
\end{equation}
Straightforwardly, if we use the same procedure in subsection \ref{quadraticregime} for both fields, we can simulate their particle productions.  To obtain the numerical result, we can set that $\psi_1$ and $\psi_2$ have coupling strength $g_1>g_2$ to show that $\psi_1$ growth should be higher than $\psi_2$. Also, to show the effect of the ceased-out mechanism, we set $\alpha_1>\alpha_2$. For the numerical Mathieu-like equation, we have $A\propto k_\psi$. It can be written for corresponding $\psi_1$ and $\psi_2$ fields as

\begin{equation}
    A_1=\Tilde{A}_1 e^{\alpha_1 t} \hspace{1cm} A_2=\Tilde{A}_2 e^{\alpha_2 t},
\end{equation}
where $\Tilde{A}_1$ and $\Tilde{A}_2$ are constants.
The result of the numerical calculation can be seen in the FIG. \ref{charmomentum}. 

Originally, if we consider fixed momentum for both fields, the growth of $\psi_1$ should be stronger than $\psi_2$. However, due to the greater suppression from the characteristic momentum $\alpha_1$ than $\alpha_2$, the $\psi_1$ production is smaller than $\psi_2$. This mechanism could select appropriate large coupling fields to be more suppressed, allowing the lower-valued coupling fields to be more abundant. Also,
this ceased-out mechanism could guarantee the appearance of the second-level preheating by suppressing the efficiency of the inflaton zero crossing. Note that this suppression by characteristic momentum is not really significant for the quartic regime since its production is already suppressed

\subsection{Tachyonic preheating}\label{tachyonpreheating}
In this part, we will study the tachyonic preheating. For this purpose, we consider the interaction part of Eq. \eqref{interaction} which can be rewritten as
\begin{equation}
    -\frac{1}{4}g_A\psi^\dagger \psi \phi^2-\frac{1}{4}g_B|\psi^\dagger|^2\phi^2-\frac{1}{4}g_C|\psi|^2 \phi^2,
\end{equation}
where we have defined the complex field $\psi=\frac{1}{\sqrt{2}}\left(\psi_a+i\psi_b \right)$, in which both $\psi_a$ and $\psi_b$ are assumed to be real fields. For simplification, we assume that the couplings in those interactions have the same strength so that we may set the coupling as $g_A= g_B = g_C = g$. This leads to the following term
\begin{equation}
    +\frac{1}{4}g\psi_b^2\phi^2.
\end{equation}
We omit the discussion of $\psi_a$ as it has no significant effect compared to $\psi_b$.

In the Einstein frame (up to the leading order), the interaction term can be written as
\begin{equation}
    +\frac{1}{2}g' \psi_b^2|\chi|,
\end{equation}
and the equation of motion in the Heisenberg representation is 
\begin{equation}\label{eompsib}
    \frac{d^2\psi_{bk}}{dt^2}+3H\Dot{\psi}_{bk}+\left( \frac{k^2_{\psi_b}}{a^2}-g' |\Tilde{\chi}\sin(mt)|\right)\psi_{bk}=0.
\end{equation}
Note that the negative sign in Eq. \eqref{eompsib} could lead to 'serious' tachyonic preheating, as will be shown later. If we neglect the expansion of the universe we obtain
\begin{equation}
    \frac{d^2\psi_{k}}{dt^2}+\left( k^2_{\psi}-g' |\Tilde{\chi}\sin(mt)|\right)\psi_{k}=0,
\end{equation}
where the subscript $b$ has been omitted for simplicity. If we consider the production during the zero crossing ($\sin(mt) \approx mt$) we obtain 
\begin{equation}
    \frac{d^2\psi_{k}}{dq^2}+\left( p^2-q\right)\psi_{k}=0,
\end{equation}
where the definitions of $p$ and $q$ are provided in Eq. \eqref{pqK}.
During the early stage of preheating, $p^2\ll q$, the analytical calculation of the energy density can be estimated as follows:

\begin{equation}\label{tachyonicenergydensity}
    \delta \rho_\psi^\text{tach} = \int^{\sqrt{|g'\Tilde{\chi}|}}_0 \frac{d^3k_\psi}{(2\pi)^3}\sqrt{|g'\Tilde{\chi}|}e^{\pi p^2}\simeq\frac{K^2}{4\pi^2}g'\Tilde{\chi} e^{\pi\frac{g'\Tilde{\chi}}{K^2}}
\end{equation}
We select the upper bound of the integration due to the minimum value of tachyonic preheating. The result of the calculation of Eq. \eqref{tachyonicenergydensity} can be expressed as
\begin{equation}
    \delta \rho_\psi^\text{tach} =\frac{m^{2/3}}{4\pi^2}\left[g \frac{M_p}{\xi \sqrt{6}}\Tilde{\chi}\right]^{5/3} \exp \left(\pi\frac{\left[g\frac{M_p}{\xi \sqrt{6}}\Tilde{\chi}\right]^{1/3}}{m^{2/3}} \right).
\end{equation}
From the last result, if the coupling $g\gtrsim 1.6 \times 10^{-5}$  ($\xi=100$), the tachyonic instability could drain the whole inflaton's energy by its first-zero crossing. This criterion is much smaller than Eq. \eqref{singlecrossingmatterdominated}, which requires $g\gtrsim 0.02 \xi$, as expected. Note that this 'purely' tachyonic preheating could jeopardize the appearance of the second-level preheating.

During the second-level preheating, when the inflaton field value drops below $M_p/\xi$, the equation of motion can be written as
\begin{equation}\label{secondpreheattachyonic}
    \frac{d^2 \psi_{k}}{dt^2}+3H\Dot{\psi}_{k}+\left(\frac{k^2_{\psi}}{a^2}-\frac{1}{2}g{\phi}^2\right)\psi_{k}=0.
\end{equation}
Note that this equation is only valid when the tachyonic preheating in the matter-dominated era cannot drain the whole inflaton's energy.

\subsection{The grand-daughter fields production}\label{granddaughter}
So far the calculation of the preheating stage has focused solely on the inflaton's resonance to produce the daughter field. However, the daughter fields also oscillate around zero and probably produce significant grand-daughter fields. 
Let us consider the scalaron $\chi$ produced from the oscillating $\psi$. The equation of motion in the Heisenberg picture can be written as
\begin{equation}
    \frac{d^2\chi_k}{dt^2}+3H\Dot{\chi}_k+\frac{k^2}{a^2}+\frac{1}{2}g'\psi^2 =0.
\end{equation}
If we neglect the expansion of the universe we obtain
\begin{equation}
        \frac{d^2\chi_k}{dt^2}+\frac{k^2}{a^2}+\frac{1}{2}g'\psi^2=0,
\end{equation}
which can be analytically shown to be insignificant. This means we can conclude that the origin of the particles during the preheating stage is solely due to the parametric resonance of inflaton oscillation. 

\section{The primordial black hole}\label{pbh}
In this part, we will review the analytical method from Ref. \cite{green2001primordial,liddle2000super} on the Primordial Black Hole (PBH). We adopt the model from these references in our inflationary model with non-minimal coupling. We will see that the abundance constraint of PBH is aligned with the condition in the early preheating which favored non-relativistic heavy matter production\footnote{The first investigation into preheating producing primordial black holes (PBH) is presented in Ref. \cite{green2001primordial} (see also Ref. \cite{liddle2000super}) using the chaotic inflation model. One can refer 
to Ref. \cite{torres2013primordial} for the numerical calculation.}. For PBH production, it should be produced by the strong parametric resonance.  It is implied (See Eq. \eqref{az}) that
\begin{equation}\label{Q}
    Q=\frac{2g M_p}{m^2\xi \sqrt{6}}\Tilde{\chi}\gg 1 \hspace{0.5cm}\text{or}\hspace{0.5cm} g\gg \frac{\lambda}{\xi}.
\end{equation}
Note that we used the definition of $m$ from Eq. \eqref{m} to obtain the right side of Eq. \eqref{Q}. 
The power spectrum on the quantity of $x$ can be written as \cite{liddle2000cosmological}
\begin{equation}
    \mathcal{P}_x=\frac{k_x^3}{2\pi^2}\braket{|x_k|^3},
\end{equation}
where $k_x=|\textbf{k}_x|$ is the comoving wavenumber of $x$ and $x_k$ are the inverse Fourier transform of $x$. The effect of the preheating in the $\psi$ production can be calculated as
\cite{kofman1997towards,liddle2000super,green2001primordial}
\begin{equation}
    \mathcal{P}_{\delta \psi}= \mathcal{P}_{\delta \psi}|_\text{end}\exp\left(2\mu_p m\Delta t\right),
\end{equation}
where $\Delta t$ is evaluated after the end of inflation. The characteristic exponent $\mu_p$ can be estimated (see Eq. \eqref{characteristicexponent} by assumed $\theta_p\simeq 0$) as
\begin{equation}
    \mu_p\simeq\frac{1}{2\pi}\ln\left( 1+2 e^{-\pi p^2}\right),
\end{equation}
where $p$ is taken from Eq. \eqref{pqK}. Thus
\begin{equation}
    p^2=\left(\frac{k_\psi}{K}\right)^2=\frac{1}{18\sqrt{Q}}\left(\frac{k_\psi}{k_\text{end}} \right)^2.
\end{equation}
We imposed $k_\text{end}$ as the comoving wavenumber of the scale which was evaluated at the exits of the Hubble radius during the end of inflation. For the strong coupling $Q\gg 1$, we obtain $\mu_p=\frac{1}{2\pi}\ln 3$.
To evaluate the power spectrum for the non-adiabatic modes, we need to evaluate the amplification of the non-adiabatic of the $\psi$ field. It is straightforward to write \cite{liddle2000super,green2001primordial}

\begin{equation}\label{pzeta}
    \mathcal{P}_{\zeta_\text{n-ad}}(k_\psi)\simeq \mathcal{A}\left( \frac{k_\psi}{k_\text{end}}\right)^3 I(p, m \Delta t),
\end{equation}
where $\zeta_\text{n-ad}$ is the non-adiabatic gauge-invariant curvature perturbation.  For solving Eq. \eqref{pzeta} it is required to solve
\begin{equation}
    \mathcal{A}=\frac{3 \times 2^{3/2}}{\pi^6 \mu_p^2}\left(\frac{\Tilde{\chi}}{M_p} \right)^2\left(\frac{H_{end}}{m}\right)^4 g^2 Q^{-1/4},
\end{equation}

\begin{equation}
    I(p, m \Delta t)\equiv \frac{3}{2}\int^{p_\text{cut}}_0 p'^2dp'\int^\pi_0 \sin\theta d\theta e^{2(\mu_{p'}-\mu_{p-p'})m\Delta t},
\end{equation}
and
\begin{equation}
    \mu_{p'}-\mu_{p-p'} \simeq -\frac{2p'\cos\theta}{2+e^{\pi p'^2}}p+\mathcal{O}(p^2),
\end{equation}
where $\theta$ is the angle between $p$ and $p'$. The constraint of the PBH is controlled by

\begin{equation}\label{beta}
    \beta=\frac{\rho_\text{pbh}}{\rho_{\text{tot}}}=\int^\infty_{\delta_c} P(\delta)d\delta,
\end{equation}
where $\rho_\text{PBH}$ and $\rho_\text{tot}$ correspond to the total energy density of PBH and universe respectively.
$P(\delta)$ corresponds to the probability distribution function. It can be expressed as
\begin{equation}
    P(\delta)=\frac{1}{\sqrt{2\pi} \sigma_\delta} \exp\left( -\frac{\delta^2}{2\sigma_\delta^2}\right),
\end{equation}
which is strongly controlled by the mass variance $\sigma_\delta$ as 
\begin{equation}\label{sigmadelta}
    \sigma_\delta^2 =\frac{16}{81}\int^\infty_0 W^2(\Tilde{k},\mathcal{R})\left( \frac{\Tilde{k}}{k}\right)^4 \mathcal{P}_{\text{n-ad}}\frac{d\Tilde{k}}{\Tilde{k}},
\end{equation}
where we used the relation \cite{green2004NewCalculation}
\begin{equation}
    \mathcal{P}_{\delta \psi}(k)= \left(\frac{2(1+w)^2}{(2+3w)^2}\right)\mathcal{P}_{\zeta_\text{n-ad}}(\Tilde{k})
\end{equation}
to obtain Eq. \eqref{sigmadelta} with $w=1/3$ for matter-dominated era during the horizon crossing ($aH=k$).
$W(\Bar{k},\mathcal{R})$ is the window function which can be depicted as
\begin{equation}
    W( \Tilde{k},\mathcal{R})=\exp \left( \frac{-\Tilde{k}^2 \mathcal{R}^2}{2}\right),
\end{equation}
where $\mathcal{R}=1/k_\psi$. 

Finally, we can calculate $\sigma_\delta$ (Eq.\eqref{sigmadelta}) to be
\begin{equation}\label{sigmafinal}
    \sigma_\delta^2\approx  \hspace{1mm} 3.6 \times 10^{10} g^{3/2}\xi^{1/2}\left(\frac{k_\psi}{M_p} \right)^3,
\end{equation}
where we used $\mathcal{P}_{\zeta_\text{n-ad}}$ under the assumption that $p_\text{cut}=1$ to  avoid the ultraviolet divergence, $m\Delta t=0$ for the beginning of preheating stage, $H\sim m$, and $\Tilde{\chi}\sim M_p$. Additionally, if we used $\delta_c=0.7$ \cite{liddle2000super,green2001primordial}, $\sigma_\delta$ from Eq. \eqref{sigmadelta} becomes the threshold values $\sigma_{\text{tresh}}\simeq 0.08$. If $\sigma_\delta>\sigma_{\text{tresh}}$, it could overproduce the PBH. Note that we used $\beta< 10^{-20}$ as our constraint.

Our calculation refers to the PBH production mainly at the beginning of preheating. Based on Eq. \eqref{sigmafinal},  appropriate values of $g$ could not be achieved if the momentum $k_\psi$ during the beginning of the preheating stage is large. Our analytical result shows that if we used $g=1$ and $\xi=100$, the appropriate order of $k_\psi$ is $\sim 10^{-4}M_p$ which is small. This result favors the discussion at the early preheating stage where the produced particle by the resonance should be heavy and non-relativistic.

\section{The Reheating Temperature}\label{reheatingtemperature}
\subsection{Reheating temperature by the Bose-Einstein condensate in the quadratic regime}\label{reheatingbose}

On the calculation of the reheating temperature by the decay of the scalaron, we will evaluate the effective decay enhanced by the Bose-Einstein condensation (BEC) as \cite{mukhanov2005physical}
\begin{equation}
    \Gamma_\text{eff}\simeq \Gamma_{\chi\rightarrow \psi\psi}(1+2\bar{n}_k),
\end{equation}
where $\Gamma_{\chi\rightarrow \psi\psi}\simeq\frac{g'^2}{32 \pi m}$
and $\bar{n}_k$ is the occupation number of $\psi$. The total energy of one $\psi$ is $m/2$. Thus, the momentum of one particle can be approximated by (see Eq. \eqref{quadraticpsi})
\begin{equation}
\begin{split}
    k_\psi (\text{max}) &\simeq \frac{m}{2} + \frac{g'}{m}\Tilde{\chi},  \\
    k_\psi (\text{min}) &\simeq \frac{m}{2} - \frac{g'}{m}\Tilde{\chi}.  \\
\end{split}
\end{equation}
With these, we obtain
\begin{equation}
    \Delta k=|k_\text{max}-k_\text{min}| = \frac{2g'}{m}\Tilde{\chi} 
\end{equation}
and the average value $k_*=m/2$.
The occupation number related to the particle density $n_\psi$ can be written as \cite{mukhanov2005physical}
\begin{equation}
    \Bar{n}_{k}=\frac{n_\psi}{V}=\frac{n_\psi}{4\pi k_*^2 \Delta k/(2\pi)^3}=\frac{ \pi^2 n_\psi}{mg' \Tilde{\chi}_\text{end}}=\frac{2\pi^2 \Tilde{\chi}_\text{end}}{g'}\frac{n_\psi}{n_{\chi_\text{end}}},
\end{equation}
where we used $n_{\chi_\text{end}}=\frac{1}{2}m\Tilde{\chi}^2_\text{end}$ which is evaluated at the end of inflation. 
Finally, the effective decay can be written by
\begin{equation}
    \Gamma_\text{eff}\simeq\frac{g'^2}{32\pi m}\left(1+\frac{4\pi^2 \Tilde{\chi}_\text{end}}{g'}\frac{n_\psi}{n_{\chi_\text{end}}} \right).
\end{equation}
The reheating is evaluated at the end of the quadratic regime which is $n_\psi/n_{\chi_\text{end}}\approx 1$ if we assume the universe is dominantly filled by $\psi$. Thus, the BEC effect will be important if we take the parameter to be ($\Tilde{\chi}_\text{end}\approx M_p$) 
\begin{equation}\label{gll}
    g\ll \xi\pi^2 4\sqrt{3}.
\end{equation}
Note that we used the definition of $g'$ from Eq. \eqref{gprime}.
As $g$ is relatively in order of smaller than $1$, The BEC is always guaranteed by Eq. \eqref{gll}, even if we take $\xi$ to be small. Thus, it is clear that the BEC effect could enhance the effective decay by $\frac{\xi \pi^2 4\sqrt{3}}{g}$ times. 
The effective decay enhanced by BEC can be written as
\begin{equation}\label{gammaeff}
    \Gamma_\text{eff} = \frac{g\pi M_p^2}{ 8\sqrt{3}\xi m},
\end{equation}
where we assumed $m_\psi \ll m$.
The reheating temperature can be written by 
\begin{equation}
   T_R=\left(\frac{90}{g^*\pi^2} \right)^{1/4}\sqrt{M_p \Gamma_\text{eff}}.
\end{equation}
By substituting Eq. \eqref{gammaeff} to the $T_R$, we obtain
\begin{equation}
    T_R \simeq 2 \times 10^{20}\sqrt{\frac{g}{\xi}}\hspace{1mm}\text{GeV},
\end{equation}
which is shown at a dangerously high temperature for efficient preheating. However, to obtain the reheating temperature by this mechanism, the inflaton should remain dominant at the end of the quadratic regime. It means that the preheating in the quadratic regime is inefficient at this time due to the violation of Eq. \eqref{singlecrossingmatterdominated}, favoring the small $g$. If the preheating in the quadratic regime is inefficient, the duration is also longer, $t_{\text{duration}}\simeq \frac{2\xi}{m}$ \cite{risdianto2022inflation}, and consequently gives a large $\xi$. Large $\xi$ is initially problematic as it jeopardizes the unitarity or the naturalness \cite{burgess2009power,burgess,barbon}. In short, reheating temperature evaluated by this mode is strongly discouraged.

\subsection{Reheating mechanism due to preheating in the quartic regime}\label{reheatingsection}

The mechanism shown in subsection \ref{reheatingbose} applies to small $g$ and large $\xi$. On the contrary, if $g$ is large, the resonance of $\psi$ production is efficient, thus the universe is dominated by $\psi$, which is heavy and non-relativistic. If the reheating temperature is obtained by the decay of $\psi$ to fermion-antifermion pairs ($f\Bar{f}$) with Yukawa coupling $Y$ via
\begin{equation}
-Y\psi \Bar{f}f,
\end{equation}
the process is extremely slow due to Pauli blocking during the quadratic regime. Simply put, fermion production by the decay of $\psi$ could occur if there is 'room' for fermions. This would be loosened during the transition between the quadratic and quartic regimes. However, this perturbative decay would not significantly affect the reheating temperature. The decay of $\psi \rightarrow \Bar{f}f$ is only effective if $\psi$ is produced during the quartic regime by inflaton oscillation, which can simultaneously decay to $\Bar{f}f$.

In our paper, we will evaluate the reheating temperature due to the first largest crossing in the quartic regime at the beginning of this regime. This method follows Ref. \cite{ballesteros2017standard, hashimoto2021inflation, risdianto2022inflation}. As mentioned earlier, the $\psi$ particles created during the early quartic regime will soon be converted to radiation by decaying into fermions. The energy density by this crossing can be evaluated by
\begin{equation}\label{deltarhobar}
    \delta \Bar{\rho}_\psi= \int^{\tau_0}_0 d\tau \Bar{\Gamma}_\psi(\tau) \Bar{m}_\psi \Bar{n}_\psi e^{-\int^\tau_0 \Bar{\Gamma}_\psi(\tau') d\tau'},
\end{equation}
where  $\tau_0=7.416$ \cite{Greene_structure}.
We defined the conformal mass 
\begin{equation}
    \Bar{m}_\psi=\sqrt{\frac{g}{2\lambda}}  \sin(c \tau),
\end{equation}
where $c=\frac{2\pi}{\tau_0}$, and the conformal number density as
\begin{equation}
    \Bar{n}_\psi=\int^\infty_0 \frac{d^3k_\psi}{(2\pi)^3}e^{-\pi \frac{k_\psi^2}{g/2\lambda}}= \frac{1}{8\pi^3}\left( \frac{g}{2\lambda}\right)^{3/2}.
\end{equation}
The $\Bar{\Gamma}_\psi$ corresponds to the conformal decay rate of the $\psi$ to the fermions $\Bar{f}f$ as
\begin{equation}
    \Bar{\Gamma}_\psi= \frac{Y^2 \Bar{m}_\psi}{32 \pi}.
\end{equation}
We assumed the fermion mass is much smaller than $\psi$.
When all requirements are met, the energy density
transferred to the fermions is averaged and estimated as

\begin{equation}
    \Bar{\rho}_\psi=2\frac{\delta\Bar{\rho}_\psi}{\tau_0}\tau
\end{equation}
In addition, 
$\delta\Bar{\rho}_\psi$ is evaluated by the Appendix \ref{bessel}. We assumed, the conformal energy $\Bar{\rho}_\psi=\frac{1}{4\lambda}$ is completely converted to radiation ($\delta \rho_\psi =\delta \rho_f$), the physical energy density can be written by
\begin{equation}
    \frac{1}{4\lambda}\left( \frac{\sqrt{\lambda} \phi_\text{crit} }{a}\right)^4=\frac{g^* \pi}{30} T_R^4.
\end{equation}
If we used $a=\frac{\phi_\text{crit}}{2\sqrt{3}M_p}\tau$, we obtain the reheating temperature as
\begin{equation}
    T_R \simeq 2 \times 10^{11}\text{GeV},
\end{equation}
where we have used $Y=10^{-6}$, $g=1$, and $\lambda=10^{-7}$. Note that the chosen large $g$ represents efficient preheating.

\begin{table}
    \centering
    \begin{tabular}{|c|c|c|c|}
    \hline
    \hline
        $\lambda$ & $g$ & $Y$ & $T_R$ (GeV)  \\
    \hline
    \hline
        $10^{-5}$ & $10^{-2}$ & $10^{-6}$&  $ 6.6 \times 10^{3}$  \\
    \hline
       $10^{-6}$ & $10^{-2}$ & $10^{-6}$&  $1.2 \times 10^{5}$  \\
    \hline
       $10^{-7}$   & $10^{-2}$ & $10^{-6}$& $2.1 \times 10^{6}$   \\
    \hline
       $10^{-8}$   & $10^{-2}$ & $10^{-6}$& $3.7 \times 10^{7}$   \\
    \hline
    \hline
        $10^{-9}$ & $1$ & $10^{-6}$& $ 6.6 \times 10^{13}$  \\
    \hline
        $10^{-9}$ & $10^{-0.5}$ & $10^{-6}$&  $ 3.7 \times 10^{12}$  \\
    \hline
       $10^{-9}$ & $10^{-1}$ & $10^{-6}$&  $2.8 \times 10^{11}$  \\
    \hline
       $10^{-9}$   & $10^{-1.5}$ & $10^{-6}$& $1.2 \times 10^{10}$   \\
    \hline
    \hline
        $10^{-9}$ & $10^{-2}$ & $10^{-5}$& $ 6.6 \times 10^{10}$  \\
    \hline
        $10^{-9}$ & $10^{-2}$ & $10^{-4}$&  $ 6.6 \times 10^{12}$  \\
    \hline
       $10^{-9}$ & $10^{-2}$ & $10^{-3}$&  $6.6 \times 10^{14}$  \\
    \hline
       $10^{-9}$   & $10^{-2}$ & $10^{-2}$& $6.5 \times 10^{16}$   \\
    \hline
    \hline
    \end{tabular}
    \caption{The reheating temperature varies with changes in the parameters $\lambda$, $g$, and $Y$. The first four results show how the temperature changes when $\lambda$ is varied while $g$ and $Y$ are fixed. The next four results show the temperature changes when $g$ is varied while $\lambda$ and $Y$ are fixed. The final four results show the temperature changes when $Y$ is varied while $g$ and $\lambda$ are fixed.}
    \label{table}
\end{table}

In Table \ref{table}, we present another example of varying $\lambda$ with fixed couplings for $Y$ and $g$. The results demonstrate that decreasing $\lambda$ by one order of magnitude could lead to a significantly higher reheating temperature $T_R$ by one to two orders of magnitude. Conversely, increasing $g$ and $Y$ by one order of magnitude could also substantially increase the reheating temperature $T_R$ by one to two orders of magnitude. It is essential to highlight that a large value for $g$ is preferred due to constraints related to primordial black hole (PBH) production, implying efficient preheating with a short duration. Note that our chosen parameters, namely $\lambda=10^{-9}$, $g=10^{-2}$, and $Y=10^{-2}$, could result in a reheating temperature on the order of $>10^{16}$ GeV, which should be disfavored \footnote{The upper bound of the reheating temperature is strongly model-dependent (see Refs. \cite{chung1999production, mazumdar2014quantifying, mcdonald2000reheating}). However, we have set it at $10^{16}$ GeV based on the reheating temperature of realistic models such as Higgs inflation \cite{Bezrukovinitials}, which is approximately $10^{15}$ GeV. Other inflationary models mostly indicate a lower temperature, so we have set one order higher than Higgs inflation as the upper bound model.} \cite{lozanov2020reheating}.

\section{Summary}\label{summary}
In this paper, we investigate inflation with non-minimal coupling. To study this inflationary model, we apply the Weyl transformation to express the action in the Einstein frame. The preheating stage after inflation is evaluated by using the Einstein frame which favors the quadratic potential $\frac{1}{2}m^2\chi^2$. We refer to this era as a quadratic regime. In our paper, this regime is also referred to as the first-level preheating.
When the inflaton drops to the critical point $\Tilde{\chi}\simeq Mp/\xi$, the Einstein and Jordan frames coincide. We refer to this condition as the second-level preheating. The potential in Eq. \eqref{potentialsplit} can be approximated as $\frac{1}{4}\lambda \chi^4$. At this point, $\Tilde{\chi}_\text{crit}\approx \Tilde{\phi}_\text{crit}$, allowing us to approximate the potential using $\phi$ as $\frac{1}{4}\lambda \phi^4$. Based on this potential, the second-level preheating is also referred to as the quartic regime. To be more practical, we use the scalaron $\chi$ as the inflaton for the quadratic regime and $\phi$ for the quartic regime. Using such conditions, we discuss the features caused by these two levels of preheating. This is the core of the findings in our paper.

After the end of inflation, the inflaton $\chi$ starts to oscillate and we enter the preheating in the quadratic regime. The energy drain of the inflaton is highest during the first several zero crossings. In our study, we have determined the parameter constraints for the scenario in which only one crossing can drain the inflaton's energy. This serves as the upper bound of our parameter constraints. However, if such a condition exists, no inflaton energy remains, meaning the second-level preheating does not occur. On the other side, during this time, the fermions production is effectively suppressed due to Pauli blocking.

After the field value drops to $M_p/\xi$, the second-level preheating begins. At this time, the potential is approximated by $\frac{1}{4}\lambda\phi^4$, which we refer to as the quartic regime due to the $\phi^4$ term in the potential. As Pauli blocking no longer exists, the production of relativistic fermions from the daughter fields becomes abundant, leading to the reheating of the universe.

We also investigated the behavior of higher-dimensional operators in both the quadratic and quartic regimes. In the quadratic regime, we show that higher-dimensional operators are still less efficient than lower-dimensional ones, but they can still show efficient behavior. Later, in the quartic regime, higher-dimensional operators are insignificant. 
Additionally, we have introduced the ceased-out mechanism, where the momentum of produced particles $\psi$ is not predetermined but follows a specific equation characterized by $\alpha$. This time-dependent momentum can suppress particle growth due to inflaton oscillation. Furthermore, the production particle can be controlled to favor the production of one species over another. 

This paper also discusses 'purely' tachyonic preheating, which can more rapidly drain the inflaton's energy than conventional preheating. It is important to note that tachyonic preheating could easily jeopardize the appearance of the second-level preheating. We also examined the production of 'granddaughter' fields during the preheating stage but found them insignificant.

We conducted a detailed analysis of primordial black hole (PBH) production during the early preheating stage, following the approach outlined in reference \cite{green2001primordial}, in the case of non-minimal coupling. We determined that PBH abundance is maximized during preheating when applying strong parametric resonance. For the parameters $\xi=100$ and $g=1$, we found that to obtain the appropriate PBH abundance, the order of momentum $\psi$ during the early stage should not exceed $10^{-4} M_p$.

Finally, we have examined the reheating temperature in two different scenarios. In the first scenario, we examined the perturbative decay of the scalaron, denoted as $\chi \rightarrow \psi \psi$, which is enhanced by the presence of BEC. The reheating temperature in this scenario is extremely high, and to tackle this, the $g/\xi$ ratio needs to be extremely small. This implies that $g$ is small and $\xi$ is large, resulting in a prolonged and inefficient preheating process. The inflaton predominantly existed after the end of the quadratic regime, and the reheating temperature was determined by its decay. In the second scenario, the reheating temperature is assessed through the decay of the daughter fields from the inflaton oscillation during the quartic regime. This scenario occurs when $g$ is large.

In the application for the realistic models (such as Higgs inflation). It is necessary to consider the Running Group Equation (RGE) on $\xi$,  $\lambda$ and $g$. But this is beyond the scope of this paper. We will consider this RGE in future work.

\section*{Acknowledgement}
We thank James M. Cline and M. Shaposhnikov for the short clarifications. We also thank Yunita K. Andriani for the small help on the numerical code.

\appendix
\section{Appendixes}\label{apx}
\subsection{The modified Bessel function of the first kind}\label{bessel}
In evaluating Eq. \eqref{deltarhobar}, we need to use the special integral as
\begin{equation}
    \int^1_0 dx  \sin^2(2 \pi x) e^{-D \sin^2(\pi x)}=\frac{2}{D}e^{-\frac{D}{2}}I_1\left(D/2\right),
\end{equation}
where
\begin{equation}
   I_v(s) =\sum^\infty_{u=0} \frac{1}{\Gamma(u+v+1)u!}\left(\frac{s}{2} \right)^{2u+v}
\end{equation}
is the modified Bessel function of the first kind.

\bibliography{apssamp}

\end{document}